\documentclass[aps]{revtex4}
\newcommand{\bea}{\begin{eqnarray}}
\newcommand{\eea}{\end{eqnarray}}
\newcommand{\ba}{\begin{array}}
\newcommand{\ea}{\end{array}}
\newcommand{\hh}{\hspace*{10mm}}

\newcommand{\h}{\hspace*{.5mm}}
\newcommand{\nn}{\nonumber}
\newcommand{\p}{\partial}
\newcommand{\s}{\star}

\newcommand{\te}{\theta}
\newcommand{\w}{\hat{\cal W}}
\newcommand{\x}{\xi}
\newcommand{\z}{\zeta}
\newcommand{\ie}{\it i.e.}
\def\br{{\bf r}}
\def\bx{{\bf x}}

\def\bk{{\bf k}}
\def\bp{{\bf p}}
\def\bq{{\bf q}}
\def\bz{{\bar z}}

\def\dx{{\rm d}{\bf x}}
\def\dk{{\rm d}{\bf k}}

\def\d{{\rm d}}
\def\la{\langle}
\def\ra{\rangle}
\def\D{\Delta}
\def\BR{{\rm\bf R}^2}

\def\bxi{\mbox{\boldmath $\xi$}}

\def\pf {\mathrm{Pf}}
\def\bte{\mbox{\boldmath $\theta$}}
\def\bo {\mbox{\boldmath $\omega$}}
\def\a{\alpha}
\def\2a{2\alpha}
\def\3a{2\alpha-1}

\begin{document}

\title{ Area Preserving Transformations in Non-commutative Space and NCCS Theory}

\author{ M. Eliashvili and G. Tsitsishvili\\
\hh\hh}
\address{Department of Theoretical Physics, A. Razmadze Mathematical Institute,
Georgian Academy of Sciences,
Aleksidze Str.1, Tbilisi 0193 Georgia\\
simi@rmi.acnet.ge}

\begin{abstract}\noindent
We propose an heuristic rule for the area transformation on
the non-commutative plane.
 The non-commutative area preserving transformations
 are  quantum deformation of the classical symplectic
diffeomorphisms. Area preservation condition is formulated as
a field equation in the non-commutative { Chern-Simons} gauge theory.
The higher dimensional generalization is suggested and the
corresponding algebraic structure - the infinite dimensional
$\sin$-Lie algebra is extracted. As an illustrative
example the second-quantized
formulation for electrons in the lowest Landau
 level is considered.

\end{abstract}

\maketitle

\section{Introduction}

In the recent papers \cite{jackiw1,jackiw2} it is raised an
intriguing question on the connection between hydrodynamics of the
incompressible fluid and the gauge field theory on the
non-commutative (NC) space. Practical realization of this idea for the
planar ($D=2$) electron system  was considered earlier in the
context of {\it Chern-Simons} (CS) description of the quantum
 {\it Hall}
effect \cite{susskind}.

Introduction of the vector potential as an  hydrodynamical
variable together with the requirement of invariance under the
classical area preserving transformations  leads to the CS gauge
theory based on the group of the symplectic transformations $Sdiff$ in
$\BR$. Non-commutative {\it Chern-Simons} (NCCS) theory is obtained
subjecting the classical symplectic structure   to the quantum
deformation.

In the present paper we propose to attribute the above
deformation of the classical algebra
 to the non-commutativity of the two-dimensional surface under consideration.
 In other terms we consider a counterpart
of area preserving diffeomorphisms (APD's)
 in the NC space and extract the corresponding symplectic structure, which,
as one may expect  turns out to be the {\it Moyal}-type deformation of
 the classical {\it Poisson} bracket.

 The non-commutative plane is represented by the pair
 of {\it Hermitian} operators $\hat x_i$ obeying
       \bea
        [\hat x_i,\hat x_k]=i\te_{ik}=
        i\te\epsilon_{ik}\hh\hh\hh (i,k=1,2) \label{eq:ccr}
        \eea
      with the constant anti-symmetric non-commutativity matrix
      $\bte$ (for the  review of NC geometry and
      adopted notations see {\it e.g.} \cite{szabo}).

In order to establish the non-commutative  analogue  of APD's
let us remind some basic definitions concerning classical symplectic structures and
APD's  \cite{jost,arnold}.

Let  $\Delta\subset\BR$ be some compact domain, described by the
{\it Cartesian} coordinates  $x_i$. The {\it Poisson} bracket is
defined by
 \bea
\{f(x),g(x)\}_P=\te_{ik}\frac{\p f}{\p x_i}\frac{\p g}{\p x_k}.
\label{pb}
\eea

 Consider a  diffeomorphism
\bea
        x_i\rightarrow x'_i=F_i(x),\hh\hh\hh \D\rightarrow \D'. \label{eq:map}
        \eea
    Under this map the  area
        \bea
        \Omega_\D=\int_\D \d^2x=\int_\D \d^2x\h\pf\bo
        \pf\{x_i,x_k\}_P\hh\hh\hh \bo=\bte^{-1}\label{eq:eq0}
\eea
changes according to the rule
\bea
        \Omega_\D\rightarrow \Omega_\D'=\int_{\D'} \d^2x'=
    \int_\D \d^2x\h\pf\bo\h \pf\{F_i,F_k\}_P. \label{eq:eq1}
        \eea
Here we use that
\bea
{\cal J}(x)=\pf\bo\h \pf\{F_i,F_k\}_P \label{eq:jac}
\eea
 is the {\it Jacobian} determinant corresponding
to the transformation (\ref{eq:map}).   The {\it Pfaffian} is defined by
 $\pf M_{ik}=(\det \h M_{ik})^{\frac{1}{2}}$.

    Infinitesimal transformations
\bea
        F_i(x)=x_i+\x_i(x) \label{eq:inf}
        \eea
      are generated by the divergenceless vector  fields $\xi_i$
    \bea
    \x_i=\te_{ik}\p_k\x, \hh\hh\hh \p_i\xi_i=0. \label{eq:apt0}
    \eea
The sought after algebraic structure can be revealed considering the
 variation (a {\it Lie} derivative) of the scalar function
 \bea
    \delta_{\bxi} f(x)=-\pounds_{\bxi}f(x) =\{\x,f\}_P .
    \eea
Generators
\bea
t[\xi]=-i\pounds_{\bxi}
 \eea
satisfy   commutation relations
 \bea
\bigg[t[\xi],t[\eta]\bigg]=t[i\{\xi,\eta\}_P] \label{eq:walg}
\eea
 which define the  {\it Lie} algebra  of  the group ${\it Sdiff}$.

In the case of $D=2N$-dimensional {\it Euclidean} space one may
 assume
that $x_i$ are canonical coordinates, {\ie}  only non-vanishing
{\it Poisson} brackets are
 \bea
  \{x_{2\alpha-1},x_{2\alpha}\}=
\te_{2\alpha-1,2\alpha}\equiv\te_\alpha>0 \hh\hh (\alpha=1,2,,,N).
\label{eq:pbN}
 \eea
  In general the canonical
coordinate system is not an  ortho-normal one,
and the constant metric tensor
$h_{ik}$ is not  diagonal. In that case under the diffeomorphism
$x_i\rightarrow F_i(x)$ the D-volume changes according to the formula

\bea
\Omega_\D=\int_\D\d^Dx\h\sqrt{\det\h h_{ik}}\h\pf\bo\pf\{x_i,x_k\}_P
\hspace*{5mm}\rightarrow\hspace*{5mm}
\Omega_\D'=\int_\D\d^Dx\h\sqrt{\det\h h_{ik}}\h\pf\bo\h\pf\{F_i,F_k\}_P. \label{eq:eq1D}
\eea

For $D>2$ divergenceless vector fields (\ref{eq:apt0}) constitute
a symplectic ({\ie} $\bte$ conserving) subgroup of the volume
preserving transformations.

\section{Area Preserving Transformations in NC {$\BR$}}

The formula (\ref{eq:eq1}) may be used with the aim to state the
area transformation rule on the NC plane. But first one has to give a
mathematical substance to the  notion of the area on the NC space.

With this purpose consider the realization of the commutation relation (\ref{eq:ccr}) in the
{\it Hilbert} space ${\cal H}$.
Operators $\hat z=\hat x_1+i\hat x_2$ and
        $\hat {\bar z}=\hat {x}_1-i\hat x_2$ satisfy the oscillator algebra
        \bea
        [\hat z,\hat \bz]=2\theta. \label{eq:fock}
        \eea
         Introduce the normalized coherent states
\bea
|\z\ra=e^{-\frac{1}{4\theta}|\z|^2}e^{\frac{1}{2\theta}\z\hat\bz}|0\ra,
\hh\hh\hh
\hat{z}|0\ra=0
\hh\hh\hh
\la\z|\z\ra=1 \label{eq:chs}
\eea
    such that
    \bea
    \hat{ z}|\z\ra=\z|\z\ra \hh\hh\hh \z=\z_1+i\z_2.
    \eea
    Here  by $\z_i$ we denote  the averages
    \bea
    \z_i=\la\z|\hat x_i|  \z\ra .\label{eq:av}
    \eea

Note that   (\ref{eq:av}) establishes the 1-1 correspondence between the coherent states
    (\ref{eq:chs}) and the points in $\BR$ \cite{perelomov,madore}
\bea
(\z_1,\z_2)\in \BR\h\h\h\h\leftrightarrow\h\h\h\h |  \z\ra\in {\cal H}.
\eea
    Using the isomorphism
    between  the domain $\D\subset \BR$ and the subspace
    ${\cal H}_\D\subset{\cal H}$,
            the area      $\Omega_\D$  can be presented
             as an integral in the $\z$-plane
\bea
      \bar\Omega_\D=\int_\D\d^2\z\h\pf\bo\h
      \la\z|\hat \Omega|\z\ra\equiv\int_\D\d^2\z=\Omega_\D
       \label{eq:are1}
      \eea
where
\bea
\hat \Omega=-\h\frac{i}{2}\h\epsilon_{ij}[\hat x_i,\hat x_j].
\eea

     The matrix element in (\ref{eq:are1})
      mimics the {\it Pfaffian}   $\pf\{x_i,x_k\}_P$
 in (\ref{eq:eq0}). Acting by analogy with (\ref{eq:eq1})
     we set, that  the operator homomorphism
     \bea
     \w[x_i]=\hat x_i\rightarrow \w[F_i]   \label{eq:qmap}
     \eea
induces the "area transformation"
      \bea
\bar\Omega_\D\rightarrow
 \bar\Omega'_{\D}=\int_\D\d^2\z\h\pf{\bo}
      \la\z|\hat \Omega'|\z\ra  \label{eq:ansatz}
      \eea
where
\bea
\hat \Omega'=-\h\frac{i}{2}\h\epsilon_{ik}\big[\w[F_i],\w[F_k]\big]=
-\h\frac{i}{2}\h\epsilon_{ik}\w[F_i\s F_k].
\eea
      Here by
      \bea
      \w[F_i]=\frac{1}{(2\pi)^2}\int \d^2p\int
    \d^2x e^{-ip_i(\hat x_i-x_i)}F_i(x  )\label{eq:Weyl}
    \eea
we denote the  symbol of the {\it Weyl} ordering and
  \bea
        f(x)\s g(x)=e^{\frac{i}{2}\te_{ik}\p_i\p'_k}
        f(x)\cdot g(x')\vert_{x'=x} \label{eq:gm}
        \eea
is the {\it Groenewold-Moyal} star product.
      It must be noted, that the area transformation
      {\it Ansatz } (\ref{eq:ansatz}) is not
      unique as well as one can use other types of operator ordering.

      Remark now, that in the formula  (\ref{eq:ansatz}) there figures an integral
                \bea
    I_\D[f]\equiv\int_\D{\d^2\z}\la\z|\w [f]|\z\ra
    \label{eq:I}
    \eea
and from  the definitions of coherent states and
{\it Weyl} symbols one easily
finds that
\bea
I_\D[f]=\int_\D\d^2\z\int \d^2x
D(\z-x)f(x),  \label{eq:qint}
\eea
where
\bea
     D(\z-x)=
\frac{1}{\pi\theta}e^{-\frac{1}{\theta}(x_i-\zeta_i)^2}.
     \eea

Thus the area transformation rule (\ref{eq:ansatz}) takes the following form
\bea
    \bar\Omega_\D\rightarrow \bar\Omega'_\D=\int_\D \d^2\z\int
     \d^2x D(\z-x)\h\pf\bo(-i)\h\pf\{F_i,F_k\}_M(x)
\label{eq:IV}
     \eea
     where
     \bea
  \{F_i,F_k\}_M(x)=
     F_i(x)\s F_k(x)-F_k(x)\s F_i(x) \label{eq:MB}
     \eea
is the  {\it Moyal} bracket.\\

Now we see, that transition to the NC case is realized
by the substitution
\bea
\{F_i,F_k\}_P(\z)\rightarrow
\{F_1,F_2\}_{NC}(\z)\equiv -i\int \d^2x
D(\z-x)\{F_i,F_k\}_M(x). \label{eq:subst }
\eea

In the commutative limit
\bea
\lim_{\te\rightarrow 0}\pf\bo\h\pf\{F_i,F_k\}_{NC}=
\lim_{\te\rightarrow 0}\pf\bo\h\{F_i,F_k\}_P={\cal J}
\eea
      is a {\it Jacobian} and the expression (\ref{eq:IV}) gives the
     classical result (\ref{eq:eq1}).
      Note that the earlier proposed  heuristic rule for the area
           transformation \cite{me} may
     be presented in the form      (\ref{eq:IV}) if one sets  $D(\z-x)= \delta^2(\z-x)$.

The requirement that   (\ref{eq:qmap})
is an area preserving  operator transformation results
\bea
\bigg[\w[F_1],\w[F_2]\bigg]=\bigg[\hat x_1,\hat x_2\bigg]=i\te.
\label{eq:V}
 \eea
The same condition may be rewritten in terms of the {\it Moyal} bracket
               \bea
           -i\pf\bo\h\pf\{F_i,F_k\}_M=1. \label{eq:Va}
     \eea
Equations (\ref{eq:V})-(\ref{eq:Va}) are  non-commutative counterparts of the
classical area preservation condition $\{F_1,F_2\}_P=\te$ and corresponding
 operator homomorphisms can be referred to as the NC or quantum APD's.

The last equations can be generalized for the $2N$-dimensional
non-commutative space defined by the commutators \bea \big[\hat
x_{2\alpha-1},\hat x_{2\alpha}\big]=i
\te_{2\alpha-1,2\alpha}\equiv i\te_\alpha \hh\hh\hh
 (\alpha=1,2,,,N).
\label{eq:cN} \eea The {\it Fock} space operators are identified by
\bea
\hat z_\alpha=\hat x_{2\alpha-1}+i\hat x_{2\alpha},
\hh\hh
\hat \bz_\alpha=\hat x_{2\alpha-1}-i\hat x_{2\alpha}
\hh\hh
\big[\hat z_\a,\hat\bz_a\big]=2\te_\a.
\eea
Define coherent states
\bea
|\z_\a\ra&=&e^{-\frac{1}{4\theta_\a}|\z_\a|^2}e^{\frac{1}{2\theta_\a}\z_\a\hat\bz_\a}|0\ra,
\hh\hh
\hat z_\a |0\ra=0
\hh\hh
\la\z_\a|\z_\a\ra=1 \label{eq:chs2}
\eea
    such that
    \bea
    \hat z_\a|\z_\a\ra=\z_\a|\z_\a\ra.
    \eea
Construct the  state vector
\bea
|\{\z\}\ra=\prod_\a\otimes |\z_\a\ra
\eea
and introduce averages
\bea
\z_i=\la\{\z\}|\hat x_i|\{\z\}\ra,\hh\hh 1\leq i\leq D.
\eea
The last relation
states the isomorphism between points
in ${\bf R}^D$ and vectors in the
    {\it Hilbert} space.

The volume transformation may be presented as
      \bea
\bar\Omega_\D\rightarrow \bar\Omega'_{\D}=
\int_\D\d^D\z\h\sqrt{\det\h h_{ik}}\h\pf{\bo}
      \la\{\z\}|\hat \Omega'|\{\z\}\ra  \label{eq:ansatzD}
      \eea
where
\bea
\hat \Omega'=\w\big[\pf_M\{F_i,F_k\}_M\big].
\eea
The star-product modified {\it Pfaffian} is  defined by
\bea
\pf_M{\bf A}=
\frac{1}{2^N N!}\h\epsilon_{i_1j_1\cdots i_Nj_N}
A_{i_1j_1}\s A_{i_1j_1}\s\cdots \s A_{i_Nj_N}
\label{eq:Vb}
\eea
where $\s$ and $ \{F_i,F_j\}_M$  correspond to the matrix $\bte$
in (\ref{eq:cN}).

The multi-dimensional analogue of the volume preservation condition (\ref{eq:Va})
 may be presented as follows
\bea
(-i)^N\pf\bo\h \pf_M\{F_{i},F_{k}\}_M=1 \label{eq:hyd}
\eea

\section{NCCS}

Consider the map
\bea
x_i\rightarrow F_i(x)=x_i+\te_{ik}a_k(x)
\hh\hh\hh (i,k=1,2,...,2N).
\eea
The basic {\it Moyal} bracket is given by
\bea
\{F_i,F_k\}_M=i\te_{ik}+i\Phi_{ik}
\eea
 where
 \bea
\Phi_{ik}&=&\te_{im}\te_{kn}[\p_ma_n-\p_na_m- i(a_m\s a_n-a_n\s a_m)]\\ \nn
\nn\\
 &\equiv&\te_{im}\te_{kn}f_{mn}.
\eea
 In terms of these
hydrodynamical variables the  volume preservation condition
(\ref{eq:hyd}) looks as follows
\bea
 \epsilon_{i_1j_1\cdots
i_Nj_N}\sum_{l=0}^{N-1} \frac{N!}{l!(N-l)!}\h\te_{i_1j_1}\cdots
\te_{i_lj_l}\Phi_{i_{l+1},j_{l+1}}\s \cdots\s \Phi_{i_N,j_N}=0.
\label{eq:NGauss}
\eea

In $D=2$
\bea
\pf[\te_{ik}+\Phi_{ik}]=\te+\frac{1}{2}\te^2\epsilon_{mn}F_{mn}
\eea
    and equation  (\ref{eq:NGauss}) takes the      form
     \bea
     \te_{ik}D_ia_k\equiv Da=
     \te_{ik}(\p_ia_k-ia_i\s a_k)=0.
     \label{eq:GL}
     \eea
One can notice the evident resemblance between (\ref{eq:GL}) and
the {\it Gau\ss}
 law in the NCCS gauge theory. This theory is described by the {\it Lagrangian}
     \bea
     {\cal L}_{NCCS}=\frac{\kappa}{2}\varepsilon^{\mu\nu\lambda}
     a_\mu \s \bigg(\p_\nu a_\lambda -
     \frac{i}{3}\{a_\nu,a_\lambda\}_M \bigg)
     \hh\hh (\mu,\nu,\lambda=0,1,2) \label{eq:lagrangian}
     \eea
     and  (\ref{eq:GL}) turns      out to be the {\it Euler-Lagrange} equation
     \bea
     \frac{\delta{\cal L}_{NCCS}}{\delta a_0}=\kappa Da=0. \label{eq:EL}
     \eea

     The infinitesimal operator transformation
     \bea
     \Delta_\lambda\w[F_i]=-i\bigg[\w[\lambda],\w[F_i]\bigg]
     =-\te_{ik}\w[\p_k\lambda+i\{\lambda,a_k\}_M]
     =-\te_{ik}\w[\delta_{gauge}a_k]\nn
     \eea
     induces the gauge transformation of  the vector field $a_k$.

In the commutative limit one recovers the  CS theory with the classical gauge
group $Sdiff$.
The corresponding non-linear symplectic CS (SCS) {\it Lagrangian} is given by \cite{manvelyan}
\bea
     {\cal L}_{SCS}=\frac{\nu}{2}\h\varepsilon^{\mu\nu\lambda}
     A_\mu  \bigg(\p_\nu A_\lambda +
     \frac{1}{3}\{A_\nu, A_\lambda\}_P \bigg)
      \label{eq:lagrangian1}
     \eea
          and gauge transformations  look as follows
         \bea
     \delta_{gauge} A_i=\p_i\lambda-\{\lambda,A_i\}_P.\label{eq:gtr1}
     \eea

 Demanding the gauge invariance under the group $Sdiff$
one arrives at the {\it Lagrangian} (\ref{eq:lagrangian1}) which could be interpreted as an
 approximation for the total NCCS {\it Lagrangian} (\ref{eq:lagrangian}).
 Transition from the SCS   to the NCCS theory
 is accomplished by the replacement $i\{f,g\}_P\rightarrow \{f,g\}_M$.
This circumstance is exploited with the goal to promote NCCS theory as an
adequate scheme for the description of  non-compressible quantum {\it Hall} fluids
 \cite{susskind}.

\section{Algebraic Structure and Electrons in LLL}

In this item we pass  to the algebraic structure
associated with the group of NC APD's and its explicit
quantum-mechanical realization.

        Consider the infinitesimal  operator transformation
    \bea
    \w[x_k]\rightarrow\w[x_k+\te_{kl}\p_l\x]=
    \hat x_k+i\bigg[\w[\x],\hat x_k\bigg] \label{eq:optr}
    \eea
    which is a non-commutative version of (\ref{eq:apt0}).

    The corresponding variation of the {\it Weyl} symbol of
    the scalar function $f(x)$ will be given by
    \bea
    \Delta_\x \w[f]= -i\bigg[ \w[\x],\w[f]\bigg]=\w[-i\{\x,f\}_M]. \label{eq:qtr}
    \eea

 Generators
 \bea
  T[\x]=\w[\x]
 \eea
obey commutation relation
        \bea
    \bigg[ T[\x], T[\eta]\bigg]=T[\{\xi,\eta\}_M] \label{eq:Walg}
     \eea
          in accord with (\ref{eq:walg}).

 The commutation relation (\ref{eq:Walg})
  describes the algebraic structure of the group of symplectic diffeomorphisms
  in the non-commutative space.
The corresponding structure constants could be fixed considering
 a special bases in the function space.  In the case of operators
\bea
 T_\bp= T[e^{i\bp \bx}] \label{eq:Tp}
\eea
the commutation relation
\bea
\big[ T_\bp, T_\bq\big]=-{2i}\sin\bigg(\frac12\h\te_{ik}p_i\h q_k\bigg) T_{\bp+\bq} \label{eq:zachos}
\eea
reproduces the well known algebra with a trigonometric structure constants
\cite{fairlie}.
 Remind, that originally the {\it Lie} brackets for the
trigonometric {\it sin}-algebra were postulated by analogy with the
{\it Virasoro}-type commutators   and corresponding structure constants were
calculated imposing  the {\it Jacobi} identities.

Application to the theory of the quantum {\it Hall} effect
is based on the assumption, that the configuration space of the system of electrons is
a NC space. In the quantum {\it Hall} states the planar system of electrons
 is exposed to  the intense orthogonal magnetic
 field ${\bf B}=(0,0,-B)$ and electrons are   constrained
  to lie in the lowest {\it Landau} level (LLL).
Non-commutative coordinates satisfy
 \bea
 [\hat r_i,\hat r_k]=-\frac{i}{B}\epsilon_{ik} \label{eq:CR}
\eea
(we use natural units $c=\hbar=1$  taking
 electron charge $e=-1$). Remind, that the commutator (\ref{eq:CR})
 arises from the {\it Dirac} bracket for the
 system with a second class constraints \cite{dirac}, \cite{jackiw}.

   For the one-particle quantum-mechanical density operator we
     take the {\it Weyl} symbol
     \bea
     \hat \rho_{QM}(\bx)&=&\w_{\br}[\delta(\bx-\br)] =
     \frac{1}{(2\pi)^2}
     \int \dk e^{-i\bk(\hat\br-\bx)}. \label{eq:QM}
     \eea
          The subscript $\br$ in (\ref{eq:QM}) means that the {\it Weyl} ordering is
     taken with respect to the operators $\hat  r_i$  satisfying  (\ref{eq:CR}).
In the same time      coordinates $ x_i$ are
     considered as  classical variables  parameterizing the plane.

Remark that
 \bea
 \int\d^2x \hat \rho_{QM}(\bx)=\w_\br[1] \eea and
\bea \int\d^2x \hat \rho_{QM}(x_i+\te_{ik}a_k(x))
=\w_\br\bigg[1-\frac{1}{2}\te_{ij}f_{ij}+\frac{1}{2}
\te_{ij}\te_{mn}(a_n\p_m f_{ij}-f_{mj}f_{ni})\bigg]+{\cal
O}(\te^3)\label{eq:SW}
\eea
in accord with the {\it
Seiberg-Witten} map \cite{seiberg},\cite{jackiw2}.

Operator (\ref{eq:QM}) obey commutation relation
\bea
\bigg[\hat \rho_{QM}(\bx'),\hat \rho_{QM}(\bx'')\bigg]=\int\d^2 x K(\bx',\bx''|\bx)
\hat \rho_{QM}(\bx).
\eea
In the kernel
\bea
K(\bx',\bx''|\bx)=\delta(\bx-\bx')\s\delta(\bx-\bx'')
-\delta(\bx-\bx'')\s\delta(\bx-\bx') \label{eq:ker}
\eea
 the  star product   is implied with respect to the  variable
 $\bx$ and the non commutativity parameter
     $\te=-1/B$.

Charge operators
\bea
\hat Q_{QM}\{\xi\}=\int \d^2x \hat \rho_{QM}(\bx)\x(\bx)=\w_\br[\xi]
\eea
generate algebra (\ref{eq:Walg})
\bea
 \bigg[\hat Q_{QM}\{\x\},\hat Q_{QM}\{\eta\}\bigg]
 =\hat Q_{QM}\{\{\xi,\eta\}_M\}.  \label{eq:QQ}
 \eea

Up to now our consideration was restricted to the one-particle
quantum mechanics and it would be instructive to develop corresponding
field theory setup.

Introduce operators
\bea
\hat b=\sqrt{\frac{B}{2}}(\hat r_1-i\hat r_2), \hh\hh \hat b^+=\sqrt{\frac{B}{2}}(\hat r_1+i\hat r_2)
\hh\hh
\big[\hat b,\hat b^+\big]=1
\eea
oscillator
\bea
|n\ra=\frac{1}{\sqrt n!}\hat b^{+n}|0\ra,\hh\hh\hh \hat b|0\ra=0 \label{eq:osc}
\eea
and coherent states
\bea
\la z|=\la 0|e^{\sqrt{\frac{B}{2}}\h\h\hat b}e^{-\frac{B}{4}|z|^2}. \label{eq:coh}
\eea

The LLL second quantized  field is given by
\bea
\hat
\psi(\bx)=\sum_{n=0}^\infty \hat f_n u_n(\bx). \label{eq:psi}
\eea
Here by $\hat f_n$ we denote the {\it Fermi} - operators
satisfying
\bea
\big[\hat f_n,\hat f^+_m\big]_+=\delta_{mn}
\eea
and $u_n$ are one-particle LLL wave functions
\bea
u_n(\bx)=\la z|n\ra  \label{eq:un}
\eea
which obey the LLL condition
\bea
\bigg(\p_\bz+\frac{B}{4}z\bigg)u_n(\bx)=0\label{eq:lll}
\eea
(we adopt  the standard complex notations $z=x_1+ix_2,\h
\p_\bz=\frac{1}{2}(\p_1+i\p_2)$).

Reminding that as a one-particle density operator we use the {\it Weyl} symbol
(\ref{eq:QM})
 define  corresponding second quantized objects: density
\bea
\hat \rho(\bx)=\sum_{m,n}\la m|\rho_{QM}(\bx)|n\ra \hat f^+_m\hat f_n
=\int\dx' \int\dx''\hat \psi^+(\bx')\la z'|\hat\rho_{QM}(\bx)|z''\ra \hat\psi(\bx'')
\label{eq:Rho}
\eea
and charge
\bea
 \hat Q\{\xi\}=\int \dx' \xi(\bx')\hat \rho(\bx'). \label{eq:Charge}
\eea
One easily verifies that
\bea
 \bigg[\hat Q\{\xi\}, \hat Q\{\eta\}\bigg]=
  \hat Q\{\{\xi,\eta\}_M\}.
    \eea
Operator transformation
\bea
\Delta_\xi \hat r_k=-i\bigg[\w_\br[\xi], \hat r_k\bigg]
\eea
induces the transformation  of the oscillator states (\ref{eq:osc})
\bea
|n\ra\rightarrow\big(1-i\w_\br[\xi]\big)|n\ra
\eea
and corresponding variation of  the matter field
\bea
\delta_\xi\hat\psi(\bx)= -i\sum_{n=0}^\infty \la z|\w_\br[\xi]|n\ra \hat f_n. \label{eq:ltr}
\eea
One easily finds, that
\bea
\bigg(\p_\bz+\frac{B}{4}z\bigg)\delta_\xi\hat\psi(\bx) =0\label{eq:lll1}
\eea
{\it i.e.} the transformation (\ref{eq:ltr}) does not violate the LLL condition.
Note, that essentially the same transformation
was used  in  \cite{iso} with the aim to establish algebra satisfied
by the LLL projected  density operators $\hat \rho^L(\bx)$.
The {\it Fourier} components of these densities obey
the commutation relation \cite{iso,stone}
\bea
\bigg[\hat \rho^{L}_{\bp},\hat \rho^{L}_{\bq}\bigg]=2i \sin
\bigg(\frac{\bp\wedge\bq}{2B}\bigg)
e^{\frac{1}{2B}\bp\cdot \bq}\hat \rho^{L}_{\bp+\bq}.\label{eq:alg1}
\eea

Note, that rescaled operators (\ref{eq:Tp})
\bea
\tilde T_\bp= e^{-\frac{\te}{4}\bp^2}T_\bp
\eea
 obey the same algebra as the operators $\hat\rho^L_\bp$.
 Different forms of commutation relations are
 related to the various possible ways of
  the operator ordering  and definitions of corresponding symbols.
  In the present paper we use  {\it Weyl} symbols
        with the  symmetric ordering of the {\it Fock} operators.
  Equally well one can apply other types of orderings and symbols ({\it e.g.} {\it Wick } normal and
anti-normal orderings)   accompanied by the appropriate
modifications of the star product.

\section{Summary}

In the present paper we introduce the notion of the finite area
on the NC plane and suggest an heuristic rule for its transformations under
the operator homomorphisms. Algebraic structure corresponding
to  APD's on
the NC plane coincides with the quantum deformation of the algebra of the group
of the classical symplectic diffeomorphisms.

Invariance under NC APD's is equivalent of the {\it Gau\ss} law
in the NCCS theory. Otherwise speaking  area preserving
transformations in the NC space are induced by  gauge potentials
satisfying  field equations of the NCCS gauge theory. The
corresponding  gauge group corresponds to   geometric
transformations
 in  the NC space.

 APD's constitute an invariance group for the incompressible fluids like
 strongly interacting electrons  in {\it Laughlin} states.
 This symmetry is embodied in the infinite-dimensional
 algebra  generated by the LLL projected
density operators \cite{iso,cappelli}.
On the other hand the standard CS gauge theory  seems
  to be an adequate  model for the description of the quantum
   {\it Hall}
 effect (see {\it e.g.} \cite{ezawa}).
In the present paper we argue, that the CS
and the infinite symmetry approaches can be unified
in the framework of the NCCS theory, where the gauge symmetry has the geometric origin.

\section{Acknowledgments}

Authors acknowledge helpful discussions with Z.Giunashvili. M.E. is  indebted to
 P.Sorba for reading the early version of the manuscript and stimulating critical remarks.

This work is in part supported by grants INTAS  00-00561 and SCOPES 7GEPJ62379.


\begin{thebibliography}{99}
\bibitem{jackiw1}
R.Jackiw, {\it Noncommuting fields and non-Abelian fluids}, arXiv:
hep-th/0305027.
\bibitem{jackiw2}
R.Jackiw, S.-Y. Pi and A.Polychronakos,
 {\it Ann. Phys. (NY)} {\bf 301}, (2002), 157.
 \bibitem{susskind}
L.Susskind, {\it The Quantum
 Hall Fluid and Non-Commutative Chern-Simons Field Theory}, arXiv: hep-th/0101029.
\bibitem{szabo}
R.J.Szabo, {\it Phys. Rep.} {\bf 378}, (2003), 207.
\bibitem{jost}
R.Jost, {\it Rev. Mod. Phys} {\bf 36}, (1964), 572.
\bibitem{arnold}
V.I.Arnold, {\it Mathematical Methods of Classical Mechanics}, (Springer-Verlag, Berlin, 1978).
\bibitem{perelomov}
A.M.Perelomov, {\it Generalized Coherent States and Their Applications}, (Springer-Verlag, Berlin, 1986).
\bibitem{madore}
J.Madore, {\it An Introduction to Noncommutative  Geometriy and its Physical Applications},
(Cambridge University Press, 1999).
\bibitem{me}
M.Eliashvili and G.Tsitsishvili, {\it Int. J. Mod. Phys.} {\bf B16}, (2002).
 3725.
 \bibitem{manvelyan}
R.Manvelyan and R.Mkrtchyan, {\it Phys.Lett.} {\bf B327}, 47 (1994).
\bibitem{fairlie}
D.B.Fairlie, P.Fletcher and C.K.Zachos, {\it Phys. Lett.}, {\bf B218},
(1989), 203.
\bibitem{dirac}
  P. Dirac, {\textit{Lectures on Quantum Mechanics}} (Belfer
Graduate School of Science, Yeshiva University, New York, 1964).
\bibitem{jackiw}
R.Jackiw, {\it Nucl.  Phys. Proc. Suppl}. {\bf 108}, (2002), 30.
\bibitem{seiberg}
N.Seiberg and E.Witten, {\it JHEP} {\bf 9909}, (1999), 032.
\bibitem{iso}
S.Iso, D.Karabali and B.Sakita, {\it Phys.  Lett.} {\bf B196}, (1992), 142.
\bibitem{stone}
J.Martinez and M.Stone, {\it Int. J. Mod. Phys.} {\bf B7}, (1993)), 4389.
\bibitem{cappelli}
A.Cappelli, C.Trugenberger and G.Zemba, {\it Nucl. Phys.} {\bf B396}, (1993), 465.
\bibitem{ezawa}
Z.F.Ezawa, {\it Quantum Hall Effects}, (World Scientific, Singapore,
2000).
     \end{thebibliography}
\end{document}